\documentclass[12pt]{iopart}
\usepackage[dvips]{graphicx}
\usepackage{isolatin1}
\usepackage{bbm}
\usepackage{psfrag}
\usepackage{iopams}

\newcommand{\beq}{\begin{equation}} 
\newcommand{\eeq}{\end{equation}}
\newcommand{\bea}{\begin{eqnarray}} 
\newcommand{\eea}{\end{eqnarray}} 

\newcommand{\ee}{{\mathrm e}} 
\newcommand{\RR}{{\mathbbm R}} 
\newcommand{\Fhyper}{{_2F_1\,}} 

\renewcommand{\labelenumi}{(\roman{enumi})}
 
\begin{document} 
 
\title{Partial equivalence of statistical ensembles and kinetic energy} 

\author{Lapo Casetti$^1$ and Michael Kastner$^2$}
\address{$^1$ Dipartimento di Fisica and Centro per lo Studio delle Dinamiche Complesse (CSDC), Universit\`a di 
Firenze, and Istituto Nazionale di Fisica Nucleare (INFN), Sezione di Firenze, via G.~Sansone 1, 50019 Sesto Fiorentino (FI), Italy}
\address{$^2$ Physikalisches Institut, Universit\"at Bayreuth, 95440 Bayreuth, Germany}
\ead{\mailto{lapo.casetti@unifi.it}, \mailto{Michael.Kastner@uni-bayreuth.de}}
 
\begin{abstract} The phenomenon of partial equivalence of statistical ensembles is il\-lus\-trat\-ed by discussing two examples, the mean-field $XY$ and the mean-field spherical model. The configurational parts of these systems exhibit partial equivalence of the microcanonical and the canonical ensemble. Furthermore, the configurational microcanonical entropy is a smooth function, whereas a nonanalytic point of the configurational free energy indicates the presence of a phase transition in the canonical ensemble. In the presence of a standard kinetic energy contribution, partial equivalence is removed and a nonanalyticity arises also microcanonically. Hence in contrast to the common belief, kinetic energy, even though a quadratic form in the momenta, has a non-trivial effect on the thermodynamic behaviour. As a by-product we present the microcanonical solution of the mean-field spherical model with kinetic energy for finite and infinite system sizes.
\end{abstract} 
 
 

\section{Introduction}
Different statistical ensembles, like the microcanonical or the canonical ones, describe different physical situations. Under certain conditions, statistical averages calculated in different ensembles are the same in the thermodynamic limit, and this situation is referred to as ensemble equivalence. Contrary to the widespread belief among physicists, ensemble equivalence does not hold in general. Prominent counterexamples are systems with long-range forces, like gravitational or unscreened electrostatic interactions.
 
An in some sense intermediate case between equivalence and nonequivalence is the only recently described partial equivalence of ensembles. When two ensembles are properly nonequivalent, there is a set of values of the control parameter in one ensemble that does not correspond to any value of the control parameter in the other ensemble. To give an example, there may be forbidden energy values, i.\,e., energy values for which a microcanonical description is perfectly legitimate, but which cannot be realized as canonical equilibrium average energies for any value of the temperature. In a situation of partial equivalence, a whole set of values of the control parameter in one ensemble corresponds to a single value of the parameter in the other ensemble: the same average energy value of the energy is obtained for a whole set of temperatures, or vice versa. 

In the present paper, we focus on a particular kind of partial equivalence between the canonical and the microcanonical ensemble, where an interval of temperatures corresponds to a single value of the energy.  Only a few examples of this kind of partial equivalence have been discussed in the literature so far. A crucial ingredient of all examples known is a bounded above Hamiltonian function, and for this reason the absence of a standard kinetic energy term, i.\,e., of a quadratic form in the momenta, is mandatory. The main goal of the present article is to discuss the phenomenon of partial equivalence as well as the way full equivalence of ensembles is restored by adding a kinetic energy term to the Hamiltonian function of the system. We observe that adding such a standard kinetic energy may have a nontrivial effect on the thermodynamic behaviour of the system.

The paper is organized as follows. After recalling in section~\ref{sec_equiv}  the definitions of nonequivalence, partial equivalence, and full equivalence of ensembles, we discuss two examples, the mean-field $XY$ and the mean-field spherical model, in sections~\ref{sec_examples} and \ref{sec_kinetic}. Considering the potential energy of these models alone, both systems exhibit partial equivalence of the microcanonical and the canonical ensembles. Furthermore, the configurational microcanonical entropy is a smooth function, whereas a nonanalytic point of the configurational free energy indicates the presence of a phase transition in the canonical ensemble. Adding a standard kinetic energy term to the Hamiltonian, as shown in section~\ref{sec_kinetic}, equivalence of ensembles is observed and phase transitions are indicated by the presence of nonanalytic points in both, the microcanonical entropy and the canonical free energy. This result is at variance with the common belief that a standard kinetic energy yields only a trivial term in the thermodynamic potentials. We conclude with a discussion of the results and their implications in section~\ref{sec_discussion}. In \ref{app_field} we supplement these results by pointing out that not only the addition of a kinetic energy term, but also a coupling to an external magnetic field may remove partial equivalence of the microcanonical and the canonical ensemble. In \ref{app_sphkin} we report the exact microcanonical solution, at finite and infinite $N$, of the mean-field spherical model with a kinetic energy term.

\section{(Non)equivalence and partial equivalence of statistical ensembles}

\label{sec_equiv}
The various statistical ensembles model different physical situations: The mic\-ro\-can\-on\-i\-cal ensemble describes an energetically isolated system with a given number of particles, confined to a fixed volume. The canonical ensemble models a system in thermal equilibrium, i.\,e., exchanging energy with the environment. The grand-canonical ensemble is the appropriate choice when a system is in thermal equilibrium and the number of particles is not fixed. These are the ensembles most widely used, but one may also consider further statistical ensembles, like those where pressure is held constant instead of volume, and many more. In general, the ensemble averages of a given observable yield different values, depending on which ensemble is used. However, in the thermodynamic limit (in case it exists), averages from different ensembles may coincide, and in this case one speaks of ensemble equivalence. Under suitable conditions on the interparticle interactions, ensemble equivalence is guaranteed to hold \cite{Ruelle}. In physical terms, these conditions are satisfied by short-ranged, bounded below potentials with a hard- or soft-core short-distance repulsion describing intermolecular forces, like the Lennard-Jones potential. In condensed matter systems (which are the common objective of statistical physics) these conditions are typically satisfied, and for this reason ensemble equivalence is taken for granted in many treatises on statistical physics.

In recent years, however, nonequivalence of ensembles has attracted a significant amount of attention. What has been known among astronomers for almost a century, i.\,e., that long-range forces like gravitational or unscreened electrostatic interactions yield different statistical averages when treated canonically or microcanonically, has been realized only recently within the statistical mechanics community. For such long-range systems, however, the energy is a nonextensive quantity, and a thermodynamic limit in the usual sense does not exist. More recently it became clear that also in systems where the energy is extensive and a proper thermodynamic limit exists, one can find ensemble nonequivalence. The main property of these systems is that the energy, though extensive, is {\em nonadditive}\/ \cite{Touchette:02}, usually due to the long-range nature of the interactions. The simplest examples of systems showing nonequivalence of ensembles are spin models with long-range interactions \cite{DauxHoldRuf:00,BarreMukamelRuffo,IsCo:01,AnRuTor:02,ElTouTu:04,BouBa:05}, other examples include self-gravitating systems \cite{LynWood:68,Lynden-Bell:99,Chavanis:06} and models of plasmas \cite{SmithONeil:90}.

As far as the canonical and the microcanonical ensemble are concerned, a comprehensive theory of equivalence and nonequivalence has been worked out (see \cite{Touchette,TouElTur:04,ElTouTu:04} for a short account and \cite{ElHaTur:00} for an extensive treatment). In the following we will restrict the discussion to the thermodynamic limit of systems in the microcanonical and in the canonical ensemble. Before entering this subject let us recall some basic definitions and fix some notation.

\subsection{Thermodynamic functions}
\label{thermodynamics}
We consider Hamiltonian systems with $N$ degrees of freedom, characterized by a Hamiltonian function ${\cal H}:\Lambda_N\to\RR$ of the form
\beq
{\mathcal H}(p,q)=\frac{1}{2}\sum_{n=1}^N p_n^2 + V(q)
\eeq
with some potential energy $V:\Gamma_N\to\RR$. We denote by $\Gamma_N$ the $N$-dimensional configuration space and by $\Lambda_N$ its cotangent bundle, i.\,e., the phase space. We shall usually denote configurations as $q=(q_1,\ldots,q_N)\in\Gamma_N$ and phase space points as $(p,q)=(p_1,\ldots,p_N,q_1,\ldots,q_N)\in\Lambda_N$. 

The fundamental quantity of the microcanonical ensemble is the {\em Boltzmann entropy}\/ or {\em microcanonical entropy}\/ as a function of the energy (per degree of freedom) $\varepsilon$,
\begin{equation}
s(\varepsilon)=\lim_{N\to\infty}\frac{1}{N}\ln \int_{\Lambda_N} \rmd p\, \rmd q\, \delta\left[{\cal H}(p,q)-N\varepsilon\right],
\end{equation}
where $\delta$ denotes the Dirac distribution.\footnote{We define all thermodynamic functions {\em per degree of freedom}, which accounts for the factor $1/N$ in the definitions. The Boltzmann constant $k_{\mathrm{B}}$ is set to unity.} A related quantity is the {\em configurational microcanonical entropy}\/ as a function of the potential energy (per degree of freedom) $v$,
\begin{equation}\label{eq:s_conf}
s_{\mathrm{c}}(v)=\lim_{N\to\infty}\frac{1}{N}\ln \int_{\Gamma_N} \rmd q\, \delta\left[V(q)-Nv\right].
\end{equation}

The fundamental quantity of the canonical ensemble is the {\em canonical free energy}\/
\begin{equation}
\varphi(\beta)=-\lim_{N\to\infty}\frac{1}{N}\ln \int_{\Lambda_N} \rmd p\, \rmd q\, \ee^{-\beta {\cal H}(p,q)},
\label{phi_def}
\end{equation}
where $\beta= T^{-1}$ is the {\em inverse temperature}.
A related quantity is the {\em configurational canonical free energy}\/
\begin{equation}\label{eq:f_conf}
\varphi_{\mathrm{c}}(\beta)=-\lim_{N\to\infty}\frac{1}{N}\ln \int_{\Gamma_N} \rmd q\, \ee^{-\beta V(q)}. 
\end{equation}
Note that the quantity usually called ``canonical free energy'' in physics textbooks is $f=\varphi/\beta$ and has the dimension of energy (or temperature, since $k_{\mathrm{B}}=1$). In contrast, $\varphi$ is an adimensional quantity, typically used in the mathematical literature and in particular in the recent literature on ensemble (non)equivalence \cite{Touchette,TouElTur:04}. 

By means of Laplace's method for the asymptotic evaluation of integrals \cite{BenOrs}, $\varphi$ is found to be the Legendre-Fenchel transform $s^*$ of $s$,
\beq
\varphi(\beta) = s^*(\beta) = \inf_\varepsilon 
\left[ \beta \varepsilon - s(\varepsilon) \right]
\label{inf}
\eeq
(and analogously for the configurational quantities).

We may also use the Legendre-Fenchel transform in order to characterize the {\em concavity}\/ of the entropy. A concave function is a real function $g:I\to\RR$ defined on some interval $I$ for which the inequality
\begin{equation}
g(ax+(1-a)y)\geqslant ag(x)+(1-a)g(y)
\end{equation}
with $a\in[0,1]$ holds for all $x,y\in I$. If the inequality is strict, $g$ is called {\em strictly concave}. The {\em concave envelope}\/ of a function $s$ is defined as the smallest concave majorant of $s$. One can show that the concave envelope of a continuous function is given by its twofold Legendre-Fenchel transform \cite{Rockafellar}
\beq\label{concave_env}
s^{**}(\varepsilon) = \varphi^*(\varepsilon) = \inf_\beta 
\left[ \beta \varepsilon - \varphi(\beta) \right].
\eeq 
A concave function coincides with its concave envelope, whereas a nonconcave function does not,
\beq
\mbox{$s(\varepsilon)$ is concave}\;\Longleftrightarrow\; s(\varepsilon) = s^{**}(\varepsilon).
\eeq
This has a remarkable consequence: The free energy $\varphi$, by definition, can always be obtained from the entropy by means of a Legendre-Fenchel transform. The converse, however, is not always true, since the correct entropy $s$ can be computed as the Legendre-Fenchel transform of $\varphi$ only in case $s$ is a concave function. 

\subsection{Full equivalence, partial equivalence, and nonequivalence of ensembles}
\label{sec_partial}
(Non)equivalence of ensembles can be defined on different levels of description. Here we consider ensemble equivalence on the thermodynamic level as discussed in \cite{Touchette}, whereas a related definition can be given making use of the concept of macrostates \protect\cite{TouElTur:04,ElHaTur:00}. At the thermodynamic level one compares the energy $\varepsilon$ in the microcanonical ensemble with the average energy in the canonical ensemble, 
\begin{equation}
\varphi'(\beta)=\frac{1}{N}\left\langle{\mathcal H}\right\rangle_\beta~,
\end{equation}
where $\left\langle \cdot \right\rangle_\beta$ stands for an average over the canonical probability measure. This leads us to the following definitions:
\begin{description}
\item[Full equivalence of ensembles:] for any value of $\varepsilon$ there exists {\em exactly}\/ one value of $\beta$ such that $\varepsilon=\varphi'(\beta)$.
\item[Partial equivalence of ensembles:] for any value of $\varepsilon$ there exists a value of $\beta$ such that $\varepsilon=\varphi'(\beta)$, but the correspondence is not one-to-one.
\item[Nonequivalence of ensembles:] there exist values of $\varepsilon$ which cannot be obtained as canonical average energies for any value of $\beta$. 
\end{description}
Note that this definition of partial equivalence differs slightly from the definition used by Ellis, Haven, and Turkington in \cite{ElHaTur:00}, where only a many-to-one relation between energies and temperatures is called partial equivalence, but not the inverse situation of a one-to-many relation. From a mathematical point of view, their non-symmetric definition may appear reasonable, but from a physical point of view we regard the above definition as preferable which treats the microcanonical and the canonical ensemble in a symmetric way.

The above definitions of (non)equivalence can be rephrased in terms of so-called
supporting lines, and the same can be done for other concepts based on the
mathematics of Legendre-Fenchel transforms. Supporting lines are a geometric
interpretation of the Legendre-Fenchel transform \eref{inf}, and we will use it
to some extent implicitly in the following. A detailed account can be found in \cite{CoElTouTu:06}.

It is useful to translate the properties given in the above definitions of (non)equiv\-a\-lence into properties of the microcanonical entropy function $s$. To this aim, we distinguish several cases in the following. We restrict ourselves to the case $s'(\varepsilon)\geq 0$. First, because it is the physically interesting case of positive temperatures, and second, because allowing $s'(\varepsilon)$ to be negative would require the discussion of further cases than those listed below. Rigorous formulations and proofs of several of the following statements can be found in \cite{ElHaTur:00}.
\begin{enumerate}
\renewcommand{\labelenumi}{(\alph{enumi})}
\item If $s$ is a {\em concave} function and its derivative $s'$ takes on {\em all}\/ values from the interval $(0,+\infty)$, then equation~\eref{inf} reduces to a standard Legendre transform
\beq\label{legendre}
\varphi(\beta) = s^*(\beta)=\beta \varepsilon(\beta) -s(\varepsilon(\beta)),
\eeq
where $\varepsilon(\beta)$ is defined implicitly as the solution of the equation
\beq
s'(\varepsilon) = \beta.
\label{beta}
\eeq
Geometrically speaking, $\varepsilon(\beta)$ is given as the energy value at which a straight line with slope $\beta$ is tangent to the graph of $s$ (see figure \ref{fig_equivalence} for an illustration). From the above specified properties of the entropy (concavity of $s$ and codomain $(0,+\infty)$ of $s'$), the one-to-one correspondence of microcanonical and canonical descriptions follows and we have full equivalence of ensembles.
\begin{figure}[ht]
\center
\psfrag{s}{$s$}
\psfrag{spu}{$s'(\varepsilon_0)$}
\psfrag{u}{$\varepsilon_0$}
\psfrag{e}{$\varepsilon$}
\psfrag{phi}{$\varphi$}
\psfrag{beta}{$\beta$}
\includegraphics[width=14cm,clip=true]{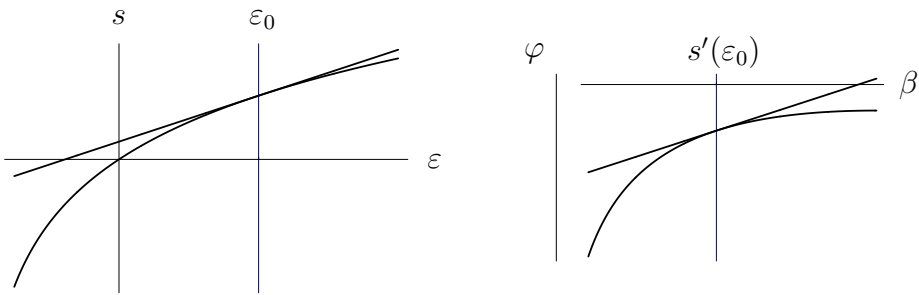}
\caption{Sketch of a case of equivalence of ensembles, illustrating the correspondence between the microcanonical entropy $s(\varepsilon)$ and  the canonical free energy $\varphi(\beta)$.}
\label{fig_equivalence}
\end{figure}

\item If $s$ is {\em not concave}, the Legendre-Fenchel transform
\beq
\varphi(\beta) = s^*(\beta) = \inf_\varepsilon 
\left[ \beta \varepsilon - s(\varepsilon) \right]
\eeq
can again be interpreted geometrically: $\varepsilon(\beta)$ is obtained as the energy value at which the topmost line with slope $\beta$ tangent to the graph of $s$ touches the graph (see figure~\ref{fig_non_partial_equivalence} (top) for an illustration). This implies that, due to the nonconcavity of $s$, there is an interval $(\varepsilon_1,\varepsilon_2)$ of energies which do not correspond to any value of $\beta$, and therefore the microcanonical and the canonical ensemble are nonequivalent. The interval $(\varepsilon_1,\varepsilon_2)$ is given by the values of $\varepsilon$ for which $s(\varepsilon)$ and its concave envelope $s^{**}(\varepsilon)$ disagree.

Nonconcave entropy functions may occur in systems with long-range interactions, and the nonconcave region of $s$ indicates the presence of a discontinuous phase transition in the canonical ensemble. The first solvable model exhibiting non\-equiv\-a\-lence of ensembles was proposed by Hertel and Thirring \cite{HertelThirring}; other examples are the mean-field Blume-Emery-Griffiths model \cite{BarreMukamelRuffo} and the mean-field $k$-trigonometric model for $k>2$ \cite{Angelani_etal:05}. 
\begin{figure}[ht]
\center
\psfrag{s}{$s$}
\psfrag{el}{$\varepsilon_1$}
\psfrag{eh}{$\varepsilon_2$}
\psfrag{e}{$\varepsilon$}
\psfrag{phi}{$\varphi$}
\psfrag{bc}{$\beta_{\mathrm{c}}$}
\psfrag{beta}{$\beta$}
\includegraphics[width=14cm,clip=true]{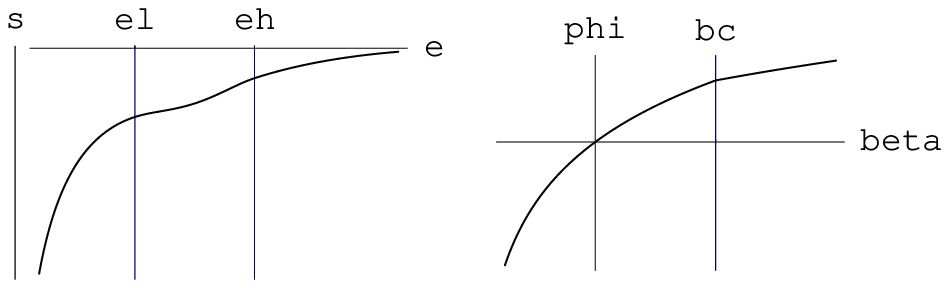}
\includegraphics[width=14cm,clip=true]{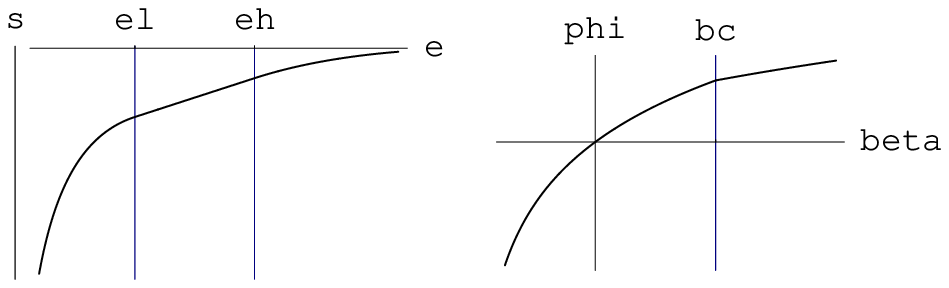}
\caption{Sketch of nonequivalence (top) due to a nonconcavity of the microcanonical entropy $s(\varepsilon)$, and of partial equivalence (bottom) due to the presence of an affine part in the entropy. In both cases the canonical free energy $\varphi(\beta)$ is singular and there is a discontinuous phase transition in the canonical ensemble.}
\label{fig_non_partial_equivalence}
\end{figure}

\item If $s$ is {\em concave, but not strictly concave}, there exists an interval of energies $(\varepsilon_1,\varepsilon_2)$ such that $s(\varepsilon)$ is affine for all $\varepsilon \in(\varepsilon_1,\varepsilon_2)$, i.\,e., its graph is a straight line in this interval (see figure~\ref{fig_non_partial_equivalence}, bottom). Then {\em all}\/ the values of the energy in that interval correspond to the {\em same} value of the inverse temperature $\beta$, equal to the slope of the affine part of $s$. Hence, although every value of $\varepsilon$ corresponds to some value of $\beta$, the correspondence is not one-to-one, and we identify this situation as partial equivalence of the microcanonical and the canonical ensemble.

An affine region of $s$ indicates the presence of a discontinuous phase transition in the canonical ensemble. In contrast to nonconcave entropy functions, such concave, but not strictly concave, entropy functions can occur for systems with short-range interactions. In fact, partial equivalence of ensembles is the typical situation observed when a discontinuous phase transition takes place in a short-range system.
\end{enumerate}

Other situations of partial equivalence of ensembles occur for {\em concave}\/ $s$ when the derivative $s'$ does not take on all values in the interval $(0,\infty)$.

\begin{enumerate}
\renewcommand{\labelenumi}{(\alph{enumi})}
\setcounter{enumi}{3}
\item Assume that $s$ is a concave function such that $s'$ takes on positive real values, but with some interval excluded,
\beq
s'(\varepsilon) \in [0,\beta_1] \cup [\beta_2,+\infty),
\eeq
with $\beta_2 > \beta_1 > 0$. This happens when the graph of $s(\varepsilon)$ has a cusp for some value $\varepsilon=\varepsilon_{\mathrm{c}}$,
\beq
s'(\varepsilon_{\mathrm{c}}^-) = \beta_2 > \beta_1 = s'(\varepsilon_{\mathrm{c}}^+)
\eeq
(see figure~\ref{fig_partial_equivalence_1} for an illustration). Then, in the microcanonical ensemble, the inverse temperature $\beta(\varepsilon)=s'(\varepsilon)$ jumps discontinuously at $\varepsilon_{\mathrm{c}}$ from $\beta_2$ to $\beta_1$, and this behaviour may be termed a {\em microcanonical discontinuous phase transition} \cite{BouBa:05,LeyvrazRuffo}. In the canonical ensemble, one obtains a free energy $\varphi(\beta)$ which is affine for $\beta \in (\beta_1,\beta_2)$: {\em all} the values of the inverse temperature in the interval $(\beta_1,\beta_2)$ correspond to the {\em same} value $\varepsilon_{\mathrm{c}}$ of the energy, and we identify this situation as partial equivalence of the microcanonical and the canonical ensemble.
\begin{figure}[ht]
\center
\psfrag{s}{$s$}
\psfrag{ec}{$\varepsilon_{\mathrm{c}}$}
\psfrag{e}{$\varepsilon$}
\psfrag{phi}{$\varphi$}
\psfrag{b1}{$\beta_1$}
\psfrag{b2}{$\beta_2$}
\psfrag{beta}{$\beta$}
\includegraphics[width=14cm,clip=true]{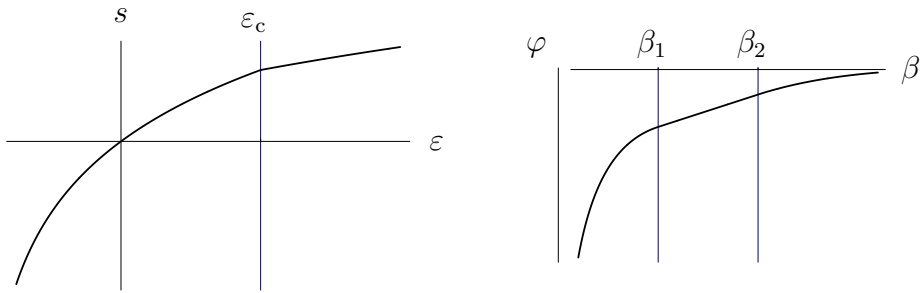}
\caption{Sketch of partial equivalence due to a cusp in the microcanonical entropy $s(\varepsilon)$. The canonical free energy $\varphi(\beta)$ has an affine part and there is a discontinuous phase transition in the microcanonical ensemble. In the canonical ensemble the first derivative of the free energy (the average energy) is continuous, but the second derivative has two discontinuities.}
\label{fig_partial_equivalence_1}
\end{figure}

We observe in figure~\ref{fig_partial_equivalence_1} that the singularity of the microcanonical entropy at $\varepsilon_{\mathrm{c}}$ corresponds to {\em two} singularities in the canonical free energy, located at $\beta_1$ and $\beta_2$. For both these inverse temperatures, the first derivative of the free energy is continuous, but the second derivative is discontinuous. As a consequence, the heat capacity is discontinuous at $\beta_1$ and $\beta_2$, and vanishes for $\beta \in (\beta_1,\beta_2)$. Therefore the {\em single}\/ discontinuous phase transition in the microcanonical ensemble corresponds to {\em two}\/ continuous phase transitions in the canonical ensemble, and these phase transitions take place at the boundaries of the region of partial equivalence. This case of partial equivalence is precisely the mirror image of case (c), with the role of the microcanonical and the canonical ensemble interchanged.

In the physics literature, several examples of systems showing discontinuous microcanonical phase transitions are discussed: gravitational systems \cite{HertelThirring,IspolatovCohen,Padmanabhan,ChavanisIspolatov}, the mean-field Blume-Emery-Griffiths model \cite{BarreMukamelRuffo}, or the mean-field $k$-trigonometric model for $k > 2$ \cite{Angelani_etal:05}. However, in all these examples---and in contrast to the graph sketched in figure~\ref{fig_partial_equivalence_1} (left)---the phase transition point is inside a ``region of nonequivalence of ensembles'', i.\,e., inside a range of energies for which the entropy $s$ does not coincide with its concave envelope. As a consequence, for the above cited systems the respective Legendre-Fenchel transform of $s$ does not show the behaviour plotted in figure~\ref{fig_partial_equivalence_1} (right).

We are not aware of any model from statistical physics properly showing the behaviour sketched in figure~\ref{fig_partial_equivalence_1}, and one may argue in favour of the claim that such a behaviour is impossible to occur. It would be a worthwhile task to make this claim more precise and to justify it rigorously.

\item As a final case, assume that the microcanonical entropy $s$ is a concave function and that $s'$ takes on values in some interval $[\beta_{\mathrm{min}},\infty)$ with $\beta_{\mathrm{min}}>0$. In particular, this situation may occur for a bounded above Hamiltonian function (see figure~\ref{fig_partial} for an illustration of this situation). The upper bound of the Hamiltonian ${\mathcal H}$ gives rise to a maximum value $\varepsilon_{\mathrm{max}}$ of the energy (per particle). Under the above stated conditions it follows from the Legendre-Fenchel transform \eref{inf} that, for all values $\beta<\beta_{\mathrm{min}}$, the canonical free energy is given by
\begin{equation}
\varphi(\beta) = \mbox{const.} + \beta \varepsilon_{\mathrm{max}}.
\end{equation}
The energy, again for inverse temperatures $\beta\in[0,\beta_{\mathrm{min}})$, is given by
\begin{equation}
\varepsilon(\beta)=\varphi'(\beta)=\varepsilon_{\mathrm{max}}.
\end{equation}
Hence, we observe that a single energy value $\varepsilon_{\mathrm{max}}$ corresponds to an interval $[0,\beta_{\mathrm{min}})$ of inverse temperatures, and we have partial equivalence of the micro\-can\-on\-i\-cal and the canonical ensemble.%
\begin{figure}[ht]
\center
\psfrag{s}{$s$}
\psfrag{em}{$\varepsilon_{\mathrm{max}}$}
\psfrag{e}{$\varepsilon$}
\psfrag{phi}{$\varphi$}
\psfrag{bc}{$\beta_{\mathrm{min}}$}
\psfrag{beta}{$\beta$}
\includegraphics[width=14cm,clip=true]{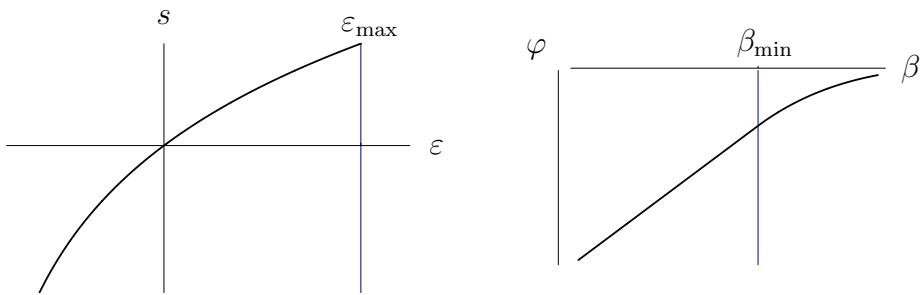}
\caption{Sketch of partial equivalence with a concave and regular microcanonical entropy. Left: microcanonical entropy $s(\varepsilon)$. Right: canonical free energy $\varphi(\beta)$.}
\label{fig_partial}
\end{figure}
This particular situation of partial equivalence has peculiar implications on the analyticity properties of the thermodynamic functions. The microcanonical entropy $s$ may be a smooth function, free of any singularities in the interior of its domain. Canonically, in contrast, a continuous phase transition at a transition inverse temperature $\beta_{\mathrm{min}}$ is signalled by a nonanalyticity in the canonical free energy.

The situation considered here is of relevance for example for certain spin systems (where the potential energy $V$ is typically bounded above and below), and in subsequent sections of this article we present two examples showing such behaviour. 
\end{enumerate}

In the above described cases (a)--(e), the role played by the canonical and the microcanonical ensembles is not completely symmetric. Nonequivalence of ensembles can only be realized as in case (b) by a nonconcave entropy. The mirror image of this situation, with the role of $s$ and $\varphi$ interchanged, is impossible to occur: since the canonical free energy $\varphi$ is given as the Legendre-Fenchel transform of $s$, it is necessarily a concave function. Similarly, the mirror image of case (e) does not exist as a consequence of the properties of the canonical free energy (see section~\ref{thermodynamics}).

From the above discussion, the intimate relationship between (non)equivalence of ensembles and phase transitions becomes clear, and for the cases (b) and (c) this has been noted a long time ago and discussed in detail \cite{HertelThirring,Touchette,TouElTur:04}. Other cases like (d) and (e) have received considerably less attention, and it is the purpose of the present article to partially fill this gap by providing examples of systems which fit into (e), and to discuss the physical implications of this type of partial equivalence.

\section{Two systems showing partial equivalence of ensembles}
\label{sec_examples}

The two model systems considered in the following are the mean-field $XY$ model and the mean-field spherical model. The interparticle interactions of these systems are of mean-field-type, i.\,e., all degrees of freedom are mutually coupled with the same strength. Considering, as is often done when studying the equilibrium statistical mechanics of spin systems, the {\em configurational}\/ thermodynamic functions \eref{eq:s_conf} and \eref{eq:f_conf} of these models, we find partial equivalence of ensembles in the sense described above. Adding a standard kinetic energy term, i.\,e., a quadratic form in the momenta, full equivalence is restored.

\subsection{Mean-field $XY$ model}
\label{sec_XY}
We start by calculating and discussing the {\em configurational}\/ entropy and free energy of the mean-field $XY$ model in this section. This model is a mean-field version of the $XY$ model of magnetism (see e.\,g.\ \cite{LeBellac}). The potential energy is given by
\beq\label{eq:potXY}
V_{XY}(\vartheta)=\frac{1}{2N}\sum_{i,j=1}^N \left[1 - \cos(\vartheta_i -\vartheta_j) \right],
\eeq
where the $N$ degrees of freedom $\vartheta_i\in [0,2\pi)$ are angular variables, yielding as configuration space an $N$-torus. In order to attain extensivity of the energy, the coupling strength between the degrees of freedom is scaled with $N$. This model has been introduced by Antoni and Ruffo \cite{AntoniRuffo} in a version with kinetic energy (which we will discuss in section~\ref{sec_kinetic}), and in the literature it is also referred to as ``Hamiltonian mean-field model''.

The potential energy $V_{XY}$ is bounded above and below: $0 \leqslant V_{XY} \leqslant \frac{N}{2}$, so that the set of the possible values of the potential energy per degree of freedom is compact, $v \in [0,\frac{1}{2}]$. The model has been solved in the thermodynamic limit---up to a maximization procedure over a single variable to be performed numerically---both in the canonical and in the microcanonical ensemble \cite{AntoniRuffo,BaBoDaRu}, and we will state some of these results in the following without derivations.

\begin{figure}[ht]
\center
\psfrag{phi}{$\varphi_{\mathrm{c}}$}
\psfrag{u}{$\!\!\langle v\rangle_{\mathrm{c}}$}
\psfrag{beta}{$\beta$}
\includegraphics[width=6.2cm,clip=true]{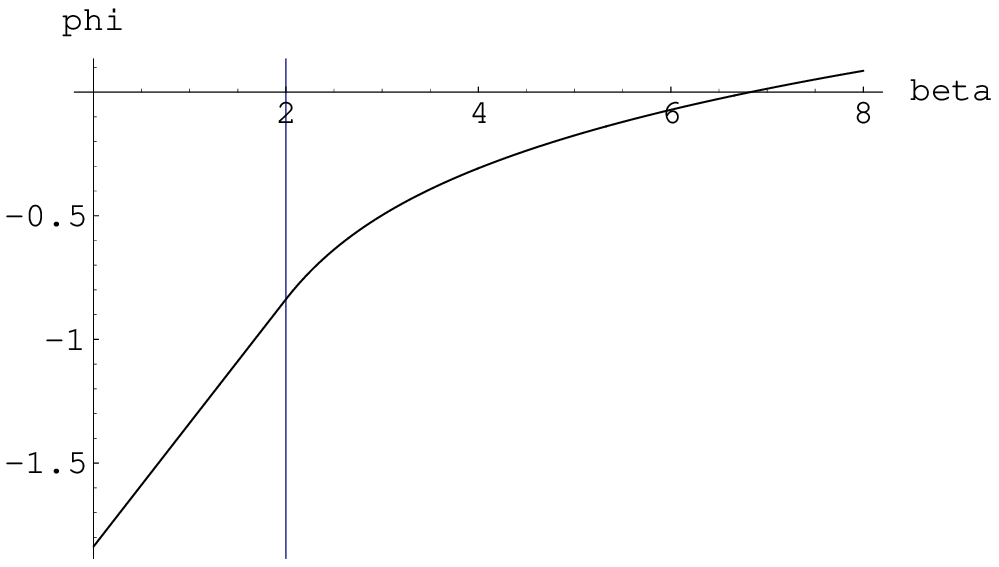}
\hspace{2mm}
\includegraphics[width=6.2cm,clip=true]{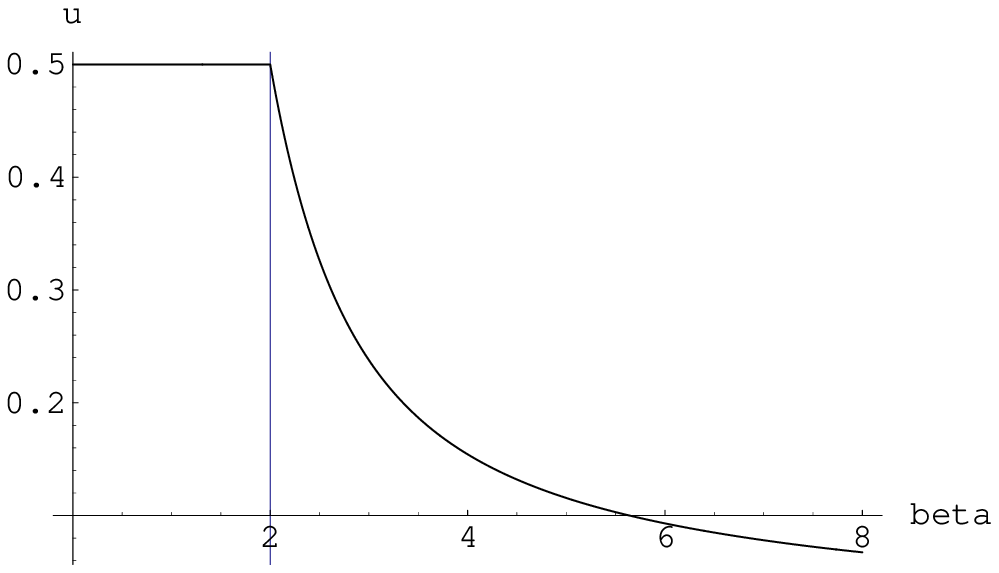}
\caption{Configurational canonical free energy $\varphi_{\mathrm{c}}(\beta)$ and canonical potential energy $\langle v \rangle_{\mathrm{c}} (\beta)$ of the mean-field $XY$ model. The vertical lines mark the critical inverse temperature $\beta=2$.}
\label{fig_free_XY}
\end{figure}

In the canonical ensemble, for the configurational free energy $\varphi(\beta)$, the expression
\beq
\varphi_{\mathrm{c}}(\beta) = \frac{\beta}{2} - \max_{y} \left[\Phi(y)\right]
\label{variational}
\eeq
is obtained, with
\beq
\Phi(y) = \log[2\pi I_0 (y)]
-\frac{y^2}{2\beta},
\eeq
where $I_0$ is the modified Bessel function of zeroth order. The maximum of $\Phi$ is found by solving the consistency equation for $y$,
\beq
\frac{y}{\beta} = \frac{I_1(y)}{I_0(y)},
\label{consist}
\eeq
where $I_1$ is the modified Bessel function of first order. As long as $\beta < 2$, the unique solution of \eref{consist} is $\tilde y = 0$, yielding a {\em linear}\/ behaviour of $\varphi_{\mathrm{c}}(\beta)$: the hallmark of partial equivalence. Numerically, for $\beta > 2$ one finds a $\beta$-dependent value $\tilde y$ as a solution of the consistency equation \eref{consist}, corresponding to the maximum in \eref{variational}. Summarizing, in the canonical ensemble we have
\beq
\varphi_{\mathrm{c}}(\beta) =
\cases{
\frac{\beta}{2} - \log (2 \pi) & \mbox{for $\beta \leqslant 2$},\\
\frac{\beta}{2} - \Phi[\tilde y (\beta)] & \mbox{for $\beta > 2$},
}
\eeq
and the result of a numerical evaluation of this expression is plotted in figure~\ref{fig_free_XY} (left).
The linear part of the graph corresponds to the region of partial equivalence: for $\beta < 2$, the average energy in the canonical ensemble is constant and equal to $\frac{1}{2}$. Differentiating $\varphi_{\mathrm{c}}$, one finds 
\beq
\langle v \rangle_{\mathrm{c}} (\beta) = \varphi_{\mathrm{c}}'(\beta) =
\cases{
{\displaystyle \frac{1}{2}} & \mbox{for $\beta \leqslant 2$}, \\
{\displaystyle \frac{1}{2}\left(1 - \frac{\tilde y(\beta)^2}{\beta^2} \right) } & \mbox{for $\beta > 2$},
}
\eeq
and a plot of this quantity can be found in figure~\ref{fig_free_XY} (right). The nonanalytic points of the configurational canonical free energy and of the canonical potential energy at $\beta = 2$ signal a continuous phase transition from an ordered phase (low temperatures) to a disordered phase (high temperatures).

\begin{figure}[ht]
\center
\psfrag{e}{$v$}
\psfrag{s}{$s_{\mathrm{c}}$}
\includegraphics[width=6.2cm,clip=true]{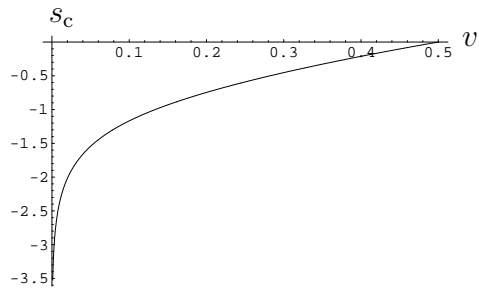}
\caption{Configurational microcanonical entropy $s_{\mathrm{c}}(v)$ of the mean-field $XY$ model.}
\label{fig_entropy_XY}
\end{figure}

A calculation of the configurational microcanonical entropy $s_{\mathrm{c}}$ by means of large deviation theory has been reported in \cite{BaBoDaRu}. As a consequence of the boundedness of $V_{XY}$, the configurational microcanonical entropy $s_{\mathrm{c}}$ is defined on the compact interval $[0,\frac{1}{2}]$, and one obtains the expression
\beq\label{sc_XY}
s_{\mathrm{c}}(v) = -\tilde\lambda(v) \sqrt{1 - 2v} + \log\big[I_0\big(\tilde\lambda(v)\big)\big],
\eeq
where $\tilde \lambda(v)$ is determined by the consistency condition
\beq
I_0\big(\tilde \lambda(v)\big)\sqrt{1 - 2v} =
I_1\big(\tilde \lambda(v)\big).
\eeq
The slope of $s_{\mathrm{c}}$ is bounded below,
\begin{equation}
\min_{\varepsilon\in[0,\frac{1}{2}]}s'_{\mathrm{c}}(\varepsilon)=s'_{\mathrm{c}}\left(\case{1}{2}\right) = 2
\end{equation}
(see figure~\ref{fig_entropy_XY} for a plot of the graph of $s_{\mathrm{c}}$). Hence the configurational microcanonical entropy displays the features of partial equivalence as specified in case (e) of section~\ref{sec_partial} and, as discussed there, $s_{\mathrm{c}}$ is a smooth function on its domain $[0,\frac{1}{2}]$.

\subsection{Mean-field spherical model}
\label{sec_spherical}
\begin{figure}[ht]
\center
\psfrag{e}{$v$}
\psfrag{s}{$s_{\mathrm{c}}$}
\includegraphics[width=6.2cm,clip=true]{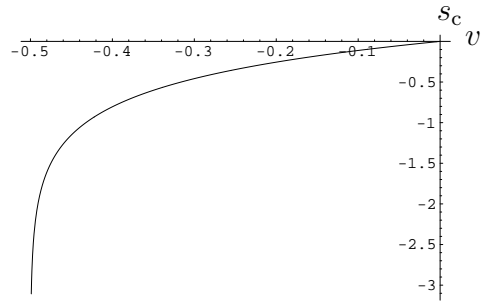}
\caption{Configurational microcanonical entropy $s_{\mathrm{c}}(v)$ of the mean-field spherical model.}
\label{fig_entropy_sph}
\end{figure}

\begin{figure}[ht]
\center
\psfrag{phi}{$\varphi_{\mathrm{c}}$}
\psfrag{u}{$\langle v \rangle_{\mathrm{c}}$}
\psfrag{beta}{$\beta$}
\includegraphics[width=6.2cm,clip=true]{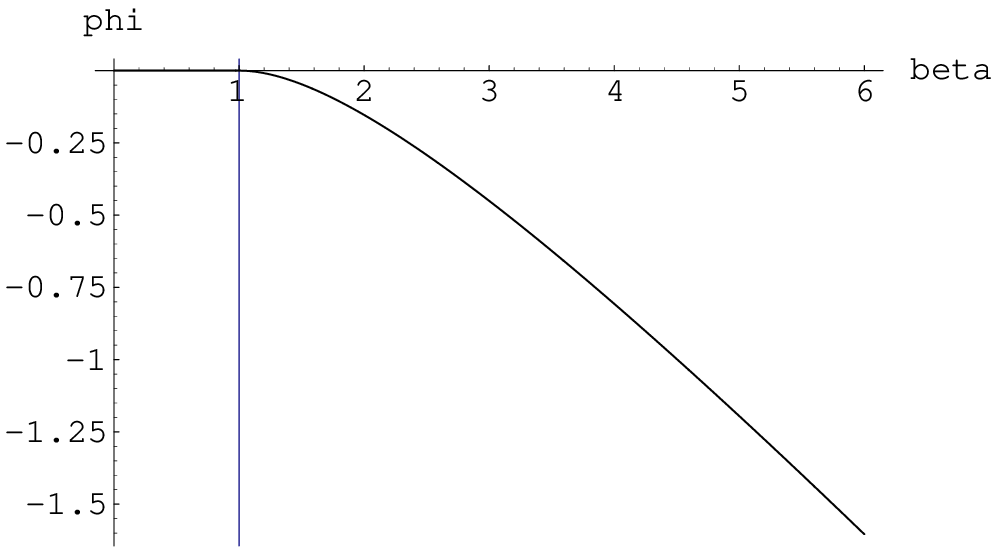}
\hspace{2mm}
\includegraphics[width=6.2cm,clip=true]{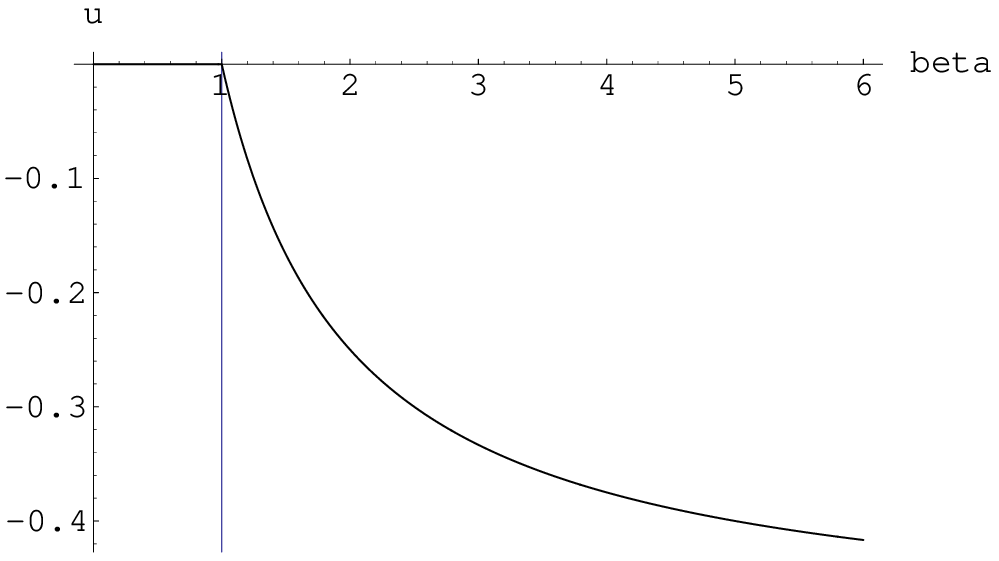}
\caption{Configurational canonical free energy $\varphi_{\mathrm{c}}(\beta)$ and canonical potential energy $\langle v \rangle_{\mathrm{c}} (\beta)$ of the mean-field spherical model.}
\label{fig_free_sph}
\end{figure}

As a second example we discuss a mean-field version of the spherical model, introduced by Kac and Berlin \cite{BerKac:52}  as a simplified and exactly solvable version of the Ising model of ferromagnetism. The potential of the mean-field spherical model is given by
\begin{equation}\label{eq:V_spherical}
V_{\mathrm{sph}}(\sigma)=-\frac{1}{2N}\sum_{i,j=1}^N \sigma_i \sigma_j,
\end{equation}
where the $N$ degrees of freedom $\sigma_i\in{\mathbbm R}$ ($i=1,...,N$) are subject to the constraint
\begin{equation}\label{eq:constraint}
\sum_{i=1}^N \sigma_i^2=N.
\end{equation}
This constraint restricts the space $\Gamma_N$ of allowed configurations $\sigma=(\sigma_1,\dots,\sigma_N)$ to an ($N$--1)-sphere with radius $\sqrt{N}$. The potential $V_{\mathrm{sph}}$ is bounded above and below, and the possible values of the potential energy per degree of freedom are $v \in \left[-\frac{1}{2},0\right]$. This model has been solved analytically for finite as well as infinite $N$ \cite{KaSchne:06}. Here we are interested in the thermodynamic limit. In this case, the configurational microcanonical entropy reads
\beq
s_{\mathrm{c}}(v) = \frac{1}{2} \log \left (1 + 2v \right),
\eeq
and the graph of $s_{\mathrm{c}}$ is plotted in figure~\ref{fig_entropy_sph}. 
$s_{\mathrm{c}}$ is a concave function on $\left[-\frac{1}{2},0\right]$, and its slope is bounded below by $s'_{\mathrm{c}}(0) = 1$. This model therefore represents another example of a system displaying partial equivalence as discussed in case (e) of section~\ref{sec_partial}. As a consequence, the configurational canonical free energy $\varphi_{\mathrm{c}}(\beta)$ is expected to be affine for $\beta \leqslant 1$, and this expectation is confirmed by the calculation of the free energy via equation \eref{inf}, yielding \cite{KaSchne:06} 
\beq
\varphi_{\mathrm{c}}(\beta) =
\cases{
0 & \mbox{for $\beta \leqslant 1$},\\
\case{1}{2}(1 - \beta +  \log \beta) & \mbox{for $\beta > 1$},
}
\eeq
(see figure~\ref{fig_free_sph} (left) for a plot of the graph of $\varphi_{\mathrm{c}}$). The canonical potential energy is
\beq
\langle v \rangle_{\mathrm{c}}(\beta) = \varphi_{\mathrm{c}}'(\beta) =
\cases{
0 & \mbox{for $\beta \leqslant 1$},\\
\frac{1 - \beta}{2\beta} &  \mbox{for $\beta > 1$}
}
\eeq
(see figure~\ref{fig_free_sph} (right) for a plot). The nonanalyticities of $\varphi_{\mathrm{c}}(\beta)$ and $\langle v \rangle_{\mathrm{c}} (\beta)$ at $\beta = 1$ signal a continuous phase transition between a low-temperature ordered (magnetized) phase ($\beta > 1$) and a high-temperature disordered phase ($\beta < 1$).

As in the case of the mean-field $XY$ model, we observe a remarkable dif\-fer\-ence between the microcanonical and the canonical ensemble: the configurational microcanonical entropy $s_{\mathrm{c}}$ is a smooth function on its domain, whereas in the canonical ensemble a phase transition is signalled by a nonanalytic point of the configurational canonical free energy.

\section{Role of the kinetic energy}
\label{sec_kinetic}

Both systems described above, the mean-field $XY$ model and the mean-field spherical model, have a continuous configuration space, and they can therefore be endowed with a standard kinetic energy $T$ which is a quadratic form in the momenta $p_i$,
\begin{equation}
T = \frac{1}{2}\sum_{i=1}^N p_i^2.
\end{equation}
In the following we will discuss the implications of adding such a standard kinetic energy term to these two models.

It is common belief that adding a standard kinetic energy term to the Hamiltonian function of some model has a trivial effect on its thermodynamic behaviour. This is correct only as long as canonical thermodynamic functions are considered: in this case, the quadratic form in the momenta results in a Gaussian integral in the canonical partition function. This integral can be easily performed, yielding a physically irrelevant additive constant in the canonical free energy $\varphi(\beta)$.

We will show in the following that, contrary to the common belief, a standard kinetic energy term may have a non-trivial effect on the thermodynamic behaviour of a system when the statistical ensembles are not fully equivalent. For the models discussed in this article, adding a kinetic energy has the effect of removing partial equivalence and of restoring full equivalence of the canonical and the microcanonical ensemble: in the presence of a kinetic energy term, the microcanonical entropy $s$ is no longer defined on a compact interval, and its slope has no positive lower bound. Furthermore, and in contrast to the configurational entropy $s_{\mathrm{c}}$, a nonanalyticity is present in the interior of the domain of $s$.

\subsection{Mean-field $XY$ model}
\label{sec_XY_kin}

\begin{figure}[ht]
\center
\psfrag{e}{$\varepsilon$}
\psfrag{s}{$s$}
\psfrag{sp}{$s'$}
\includegraphics[width=6.2cm,clip=true]{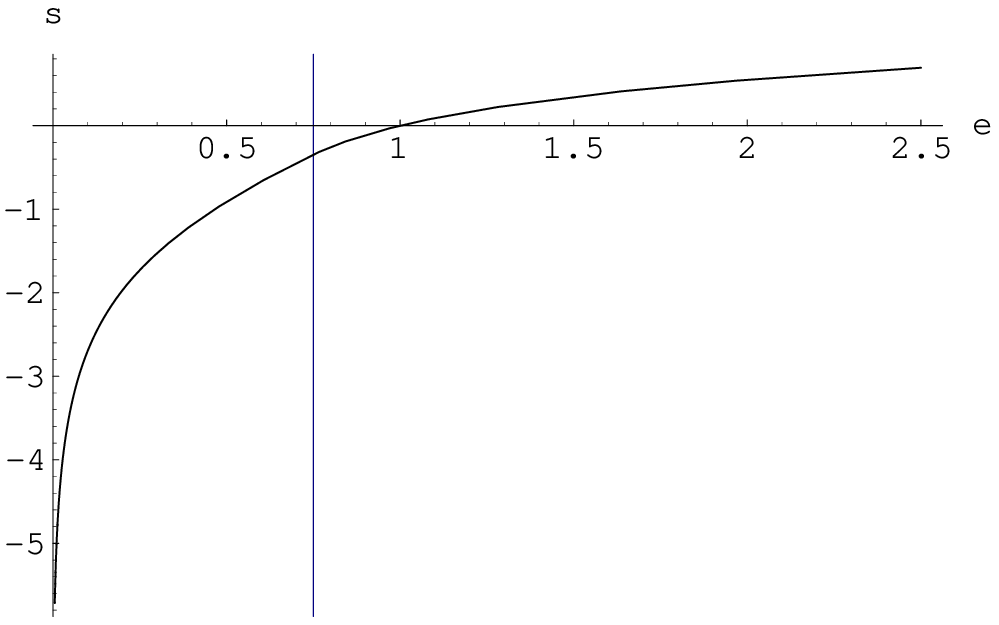}
\hspace{2mm}
\includegraphics[width=6.2cm,clip=true]{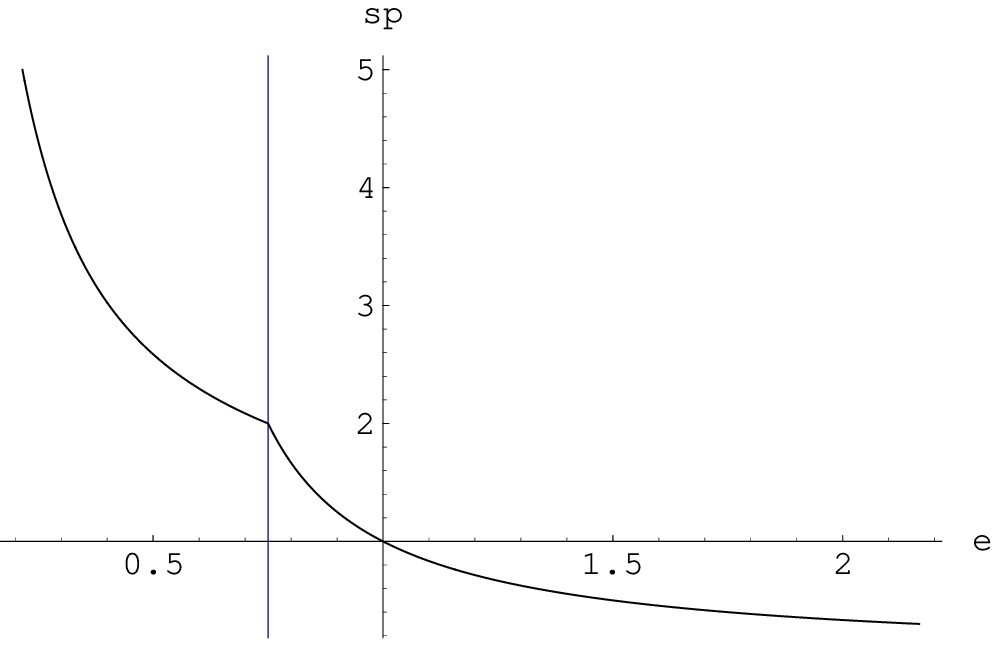}
\caption{Microcanonical entropy $s(\varepsilon)$ (left) and its derivative $s'(\varepsilon)$ (right) for the mean-field $XY$ model.}
\label{fig_dsde_XY_kin}
\end{figure}

Adding a standard kinetic energy term to the potential \eref{eq:potXY} of the mean-field $XY$ model, the Hamiltonian function
\beq
{\cal H}_{XY}(\vartheta,p_\vartheta)=\frac{1}{2}\sum_{i=1}^N 
{p_\vartheta}_i^2 + V_{XY}(\vartheta)
\label{H_XY_kin}
\eeq
is obtained, where the ${p_\vartheta}_i=\dot{\vartheta}_i$ are the momenta conjugate to $\vartheta_i$. The dynamical properties of this model have been studied in some detail, and peculiar features like quasi-stationary states were observed \cite{Yamaguchi:04}. Calculations of thermodynamic functions of this model have been reported for example in \cite{BaBoDaRu}, where by means of a large deviation computation the microcanonical entropy
\beq\label{eq:s_sup}
s(\varepsilon) = \sup_{u>0}\left[\frac{1}{2} \log u +
s_{\mathrm{c}}\left(\varepsilon - \frac{u}{2} \right)\right]
\eeq
is obtained, where $s_{\mathrm{c}}$ is the configurational microcanonical entropy \eref{sc_XY}. The supremum in \eref{eq:s_sup} is attained when
\beq\label{eq:sup_cond}
\frac{1}{u} = s'_{\mathrm{c}}\left( \varepsilon - \frac{u}{2} \right),
\eeq
and the microcanonical entropy can be written as
\beq
s(\varepsilon) = \frac{1}{2} \log \tilde u(\varepsilon) +
s_{\mathrm{c}}\left(\varepsilon - \frac{\tilde u(\varepsilon)}{2} \right),
\eeq
where $\tilde u(\varepsilon)$ is given as the solution of \eref{eq:sup_cond}. The entropy $s$ is a strictly concave function on its domain $[0,+\infty)$ and $s'$ takes on all values in $[0,+\infty)$. Therefore the microcanonical and the canonical ensembles are fully equivalent (see figure~\ref{fig_dsde_XY_kin} (left) for a plot of the graph of $s$). A phase transition is signalled by a nonanalyticity of $s(\varepsilon)$ at $\varepsilon= \frac{3}{4}$, resulting in a kink in $s'$ and in a discontinuity in $s''$ at this value of the energy (see figure~\ref{fig_dsde_XY_kin} (right) for a plot of the graph of $s'$).

\subsection{Mean-field spherical model}
\label{sec_sph_kin}
Adding a kinetic energy term to the potential $V_{\mathrm{sph}}$ of the mean-field spherical model, the Hamiltonian function
\begin{equation}\label{H_sph_kin}
{\cal H}_{\mathrm{sph}}(\sigma,p_\sigma)= \frac{1}{2}\sum_{i=1}^N {p_\sigma}_i^2
+ V_{\mathrm{sph}}(\sigma)
\end{equation}
is obtained, where the ${p_\sigma}_i=\dot{\sigma}_i$ are the momenta conjugate to $\sigma_i$. As a consequence of the spherical constraint \eref{eq:constraint}, the momenta are subject to the condition
\begin{equation}\label{eq:constraint2}
0=\frac{1}{2}\frac{\rmd}{\rmd t}\sum_{i=1}^N \sigma_i^2=\sum_{i=1}^N \sigma_i\,{p_\sigma}_i.
\end{equation}

\begin{figure}[ht]
\center
\psfrag{e}{$\varepsilon$}
\psfrag{s}{$s$}
\includegraphics[width=6.2cm,clip=true]{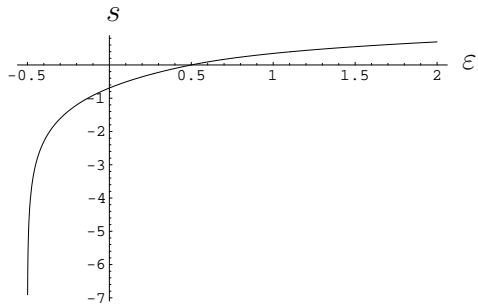}
\caption{Microcanonical entropy $s(\varepsilon)$ of the mean-field spherical model with kinetic energy term.}
\label{fig_entropy_sph_kin}
\end{figure}

For the mean-field spherical model, analytic calculations are feasible in both the microcanonical and the canonical ensemble, not only in the thermodynamic limit, but also for any finite number $N$ of degrees of freedom. The computation of the microcanonical entropy at finite and infinite $N$ is reported in \ref{app_sphkin}. In the thermodynamic limit, the microcanonical entropy can be written as
\beq\label{entropy_sph_kin}
s(\varepsilon) =
\cases{
\log\left[ \case{1}{2}(1 + 2\varepsilon)\right] &
\mbox{for $\varepsilon \leqslant \frac{1}{2}$}, \\
\case{1}{2}\log\left(2\varepsilon \right) &
\mbox{for $\varepsilon > \frac{1}{2}$}.
}
\eeq
$s$ is a strictly concave function on its domain $[-\frac{1}{2},\infty)$, and $s'$ takes on all values in $[0,+\infty)$ (see figure~\ref{fig_entropy_sph_kin} for a plot of the graph of $s$). Hence the microcanonical entropy $s$ of the mean-field spherical model with kinetic energy fulfills the conditions of case (a) in section~\ref{sec_partial}, guaranteeing full equivalence of the microcanonical and the canonical ensemble. Furthermore, $s(\varepsilon)$ has a nonanalytic point at $\varepsilon=\frac{1}{2}$, signalling the occurrence of a continuous phase transition at this value of the energy.

\section{Discussion and conclusions}
\label{sec_discussion}
We have discussed two models, the mean-field $XY$ model and the mean-field spherical model, whose configurational microcanonical entropies $s_{\mathrm{c}}$ are concave functions defined on a compact interval and whose slope $s_{\mathrm{c}}'$ has a positive lower bound. As a consequence, the micro\-can\-on\-i\-cal and the canonical ensemble are partially equivalent. Partial equivalence is removed and full equivalence is restored when, instead of configurational quantities, the ``full'' thermodynamic functions $s$ and $\varphi$ are considered. This behaviour has remarkable consequences, which are the main results of the present article and which are summarized and discussed in the following.

At variance with the common belief, the contribution of a standard kinetic energy term to the thermodynamic behaviour of a system is not always trivial. Whereas in the canonical ensemble the addition of a kinetic energy term results only in a physically irrelevant additive constant in the free energy $\varphi(\beta)$ (corresponding to a linear contribution to the ``usual'' free energy $f(T)$), the effect of such a term can be more substantial in the microcanonical ensemble. For the models discussed in the present article, we have observed the following effects when adding a kinetic energy term to the respective potential energy functions:
\begin{enumerate}
\renewcommand{\labelenumi}{(\roman{enumi})}
\item Partial equivalence of the microcanonical and the canonical ensemble is removed and full equivalence is restored.
\item The configurational microcanonical entropy $s_{\mathrm{c}}$ is a smooth function for both, the mean-field $XY$ model and the mean-field spherical model. Adding a standard kinetic energy term, a nonanalyticity shows up in the microcanonical entropies $s$, signalling the occurrence of a phase transition in these models. This indicates that care is advisable when, in order to simplify a computation, a kinetic energy term is disregarded and only the configurational part of some model is considered: in order to identify a phase transition, it is in general not sufficient to consider the analyticity properties of the configurational microcanonical entropy, but special attention has to be paid to the influence of the boundary points of its domain.
\end{enumerate}
The conclusions drawn from the two models discussed should extend to all systems with bounded above potential energy function $V$ for which the slope of the con\-fig\-u\-ra\-tion\-al microcanonical entropy $s_{\mathrm{c}}$ has a positive lower bound (case (e) of section~\ref{sec_partial}).

So for which type of models can this happen? Boundedness of potential energy functions is a widespread property among spin models, but only a few examples are known for which the slope of the configurational microcanonical entropy has a positive lower bound. To the best of our knowledge, all these examples are long-range models, but we are not aware of any argument confirming the necessity of long-range interactions to obtain such a bound of $s_{\mathrm{c}}'$. This might be an interesting point for future investigations.

\section*{Acknowledgments}
We would like to thank Oliver Schnetz for suggesting the change of coordinates employed to arrive at equation~\eref{eq:Omega2}
. L.\,C.\ thanks the Physikalisches Institut, Universit\"at Bayreuth, for hospitality, and M.\,K.\ thanks the Dipartimento di Fisica, Universit\`a di Firenze, for hospitality. L.\,C.\ and M.\,K.\ acknowledge financial support from the PRIN05-MIUR project {\em Dynamics and thermodynamics of systems with long-range interactions}. M.\,K.\ acknowledges financial support by the Deutsche Forschungsgemeinschaft (grant KA2272/2) and by INFN-Iniziativa Specifica FI61.

\appendix

\section{Mean-field $XY$ model with external magnetic field}
\label{app_field}
Partial equivalence of the microcanonical and the canonical ensemble can be removed not only by adding a kinetic energy term. In this appendix, it is shown that partial equivalence may be removed and full equivalence may be restored also by switching on an arbitrarily small external magnetic field.

\begin{figure}[ht]
\center
\psfrag{-0.4}{$-0.4$}
\psfrag{-0.2}{$-0.2$}
\psfrag{0.2}{$0.2$}
\psfrag{0.4}{$0.4$}
\psfrag{0.6}{$0.6$}
\psfrag{-0.5}{$-0.5$}
\psfrag{-1.0}{$-1.0$}
\psfrag{-1.5}{$-1.5$}
\psfrag{-2.0}{$-2.0$}
\psfrag{v}{$v$}
\psfrag{s}{$s_{\mathrm{c}}$}
\includegraphics[width=6.2cm,keepaspectratio=true]{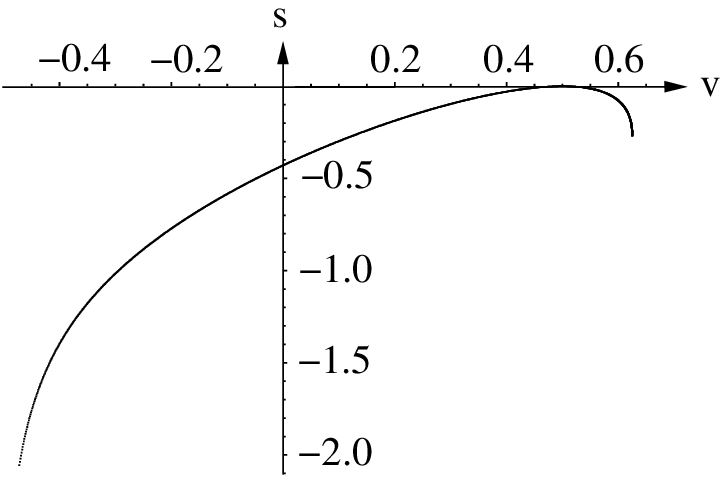}
\caption{\label{fig_entropy_XY_h}
Configurational microcanonical entropy $s_{\mathrm{c}}(v)$ of the mean-field $XY$ model with external magnetic field $h=\frac{1}{2}$. Note that the slope of $s_{\mathrm{c}}$ is unbounded above and below.}
\end{figure}

Adding to the potential \eref{eq:potXY} of the mean-field $XY$ model discussed in section~\ref{sec_XY} a coupling to an external magnetic field $h$,
\begin{equation}
V^h_{XY}(\vartheta)=\frac{1}{2N}\sum_{i,j=1}^N \left[1 - \cos(\vartheta_i -\vartheta_j) \right]-h\sum_{i=1}^N\cos\vartheta_i,
\end{equation}
the configurational microcanonical entropy can again be computed by a large deviation technique. We refrain from writing down the analytic expression for $s_{\mathrm{c}}$ in this case, but present a plot of the graph in figure~\ref{fig_entropy_XY_h}. As may be read off from the graph, the slope of $s_{\mathrm{c}}$ is not anymore bounded, and instead of partial equivalence we find full equivalence of ensembles for the mean-field $XY$ model in the presence of a non-zero external magnetic field.

Qualitatively the same happens for the mean-field spherical model: adding a coupling to an external magnetic field to the potential \eref{eq:V_spherical} of the mean-field spherical model, we obtain
\begin{equation}
V^h_{\mathrm{sph}}(\sigma)=-\frac{1}{2N}\sum_{i,j=1}^N \sigma_i \sigma_j-h\sum_{i=1}^N\sigma_i,
\end{equation}
where the $\sigma_i$ are still subject to the constraint \eref{eq:constraint}. The configurational mic\-ro\-can\-on\-i\-cal entropy for this model is given by
\begin{equation}
s_{\mathrm{c}}(v)=\frac{1}{2}\ln\left[1-\left(|h|-\sqrt{h^2-2v}\right)^2\right]
\end{equation}
(see \cite{KaSchne:06} for a derivation of this result and for a plot of the graph of $s_{\mathrm{c}}$). Like in the case of the mean-field $XY$ model, the slope of $s_{\mathrm{c}}$ has not anymore a positive lower bound, and we observe full equivalence of the microcanonical and the canonical ensemble.

\section{Mean-field spherical model in the microcanonical ensemble}
\label{app_sphkin}

The fundamental quantity of the microcanonical ensemble is the density of states as a function of the energy per degree of freedom $\varepsilon$ which, for a system with $N$ degrees of freedom characterized by a Hamiltonian function ${\mathcal H}_N:\Lambda_N\to{\mathbbm R}$, is defined as
\begin{equation}
\Omega_N(\varepsilon) \propto \int_{\Lambda_N} {\mathrm d}x\, \delta\left[{\mathcal H}_N(x)-N\varepsilon\right].
\end{equation}
We restrict ourselves to proportionalities, since multiplicative constants in the density of states result only in physically irrelevant additive constants in the microcanonical entropy $s_N=\frac{1}{N}\ln\Omega_N$. For the mean-field spherical model as defined by the Hamiltonian function \eref{H_sph_kin}, the density of states can be written as
\begin{equation}\label{eq:Omega2}
\fl\Omega_N(\varepsilon) \propto \int_{{\mathbbm R}^N}\!\!{\mathrm d}\sigma \int_{{\mathbbm R}^N}\!\!{\mathrm d}\dot{\sigma}\,\delta\!\left[\sum_{i=1}^N \sigma_i^2-N \right]\delta\!\left[\sum_{i=1}^N \sigma_i\dot{\sigma}_i\right]\delta\!\left[\frac{1}{N}\left(\sum_{i=1}^N \sigma_i \right)^2 \!- \sum_{i=1}^N \dot{\sigma}_i^2 +2N\varepsilon \right],
\end{equation}
where $\sigma=(\sigma_1,\dots,\sigma_N)^{\mathsf T}$ and $\dot{\sigma}=(\dot{\sigma}_1,\dots,\dot{\sigma}_N)^{\mathsf T}$.\footnote{To ease the notation, we use $\dot\sigma$ instead of $p_\sigma$ to denote the momenta conjugate to $\sigma$.} The first $\delta$-distribution in this expression accounts for the spherical constraint \eref{eq:constraint}, the second ensures the velocity $\dot{\sigma}$ to be tangent to the configuration space of the spherical model (constraint \eref{eq:constraint2}), and the third foliates phase space into shells of constant energy $N\varepsilon$. The form of the integral in \eref{eq:Omega2} suggests the use of $N$-dimensional spherical coordinates $r,\vartheta,\varphi,\dots$ and $\dot{r},\dot{\vartheta},\dot{\varphi},\dots$ for the $\sigma$ and the $\dot{\sigma}$ integrations, respectively. With this choice of coordinates, we have $\sum_{i=1}^N \sigma_i^2=r^2$ and $\sum_{i=1}^N \dot{\sigma}_i^2=\dot{r}^2$. Choosing the polar axis of the position coordinates $\sigma$ along the space diagonal $(1,\dots,1)$, and the polar axis of the momentum coordinates $\dot{\sigma}$ in the direction of $\sigma$, we can write $\sum_{i=1}^N \sigma_i=\sqrt{N}r\cos\vartheta$ and $\sum_{i=1}^N \sigma_i\dot{\sigma}_i=r\dot{r}\cos\dot{\vartheta}$. (Note that the first sum is the scalar product of $\sigma$ and $(1,\dots,1)$, the second sum is the scalar product of $\sigma$ and $\dot{\sigma}$.) These orientations of the polar axes render the integrand of \eref{eq:Omega2} independent of all angles but the polar angles $\vartheta$ and $\dot{\vartheta}$. Performing the trivial $2(N-2)$ other angle integrations, we obtain the proportionality 
\begin{eqnarray}
\fl\Omega_N(\varepsilon) \propto \int_0^\infty {\mathrm d}r\, r^{N-1}\,\delta\left(r^2-N\right)\int_0^\infty {\mathrm d}\dot{r}\, \dot{r}^{N-1} \int_0^\pi {\mathrm d}\vartheta\,\sin^{N-2}\vartheta \nonumber\\
\times\int_0^\pi {\mathrm d}\dot{\vartheta}\,\sin^{N-2}\dot{\vartheta}\,
\delta\left(r\dot{r}\cos\dot{\vartheta}\right)\,\delta\left(r^2\cos^2\vartheta-\dot{r}^2+2N\varepsilon\right).
\end{eqnarray}
The first and second of the $\delta$-distributions allow to easily perform the $r$- and the $\dot{\vartheta}$-integrations, yielding
\begin{equation}\label{eq:Omega4}
\Omega_N(\varepsilon) \propto \int_0^\pi \!{\mathrm d}\vartheta\,\sin^{N-2}\vartheta \int_0^\infty \!\!{\mathrm d}\dot{r}\, \dot{r}^{N-3}\,\delta\left(\dot{r}-\sqrt{N(\cos^2\vartheta+2\varepsilon)}\right).
\end{equation}
We observe that this integral can be non-vanishing only if $\cos^2\vartheta+2\varepsilon\geqslant0$, and, performing the $\dot{r}$-integration, we write
\begin{equation}\label{eq:Omega5}
\Omega_N(\varepsilon) \propto \int_0^\pi {\mathrm d}\vartheta\,\sin^{N-2}\vartheta \left(\cos^2\vartheta+2\varepsilon\right)^{(N-3)/2} \Theta\left(\cos^2\vartheta+2\varepsilon \right),
\end{equation}
where $\Theta$ denotes the Heaviside step function. With the substitution of variables $y=\cos^2\vartheta$, we obtain the expression
\begin{equation}\label{eq:Omega6}
\Omega_N(\varepsilon) \propto \int_0^1 {\mathrm d}y\,y^{-1/2}\left(1-y\right)^{(N-3)/2}\left(2\varepsilon+y\right)^{(N-3)/2} \Theta\left(2\varepsilon+y\right)
\end{equation}
for the density of states. Distinguishing the three cases
\begin{equation}
\fl\Omega_N(\varepsilon) \propto
\cases{
0 & \mbox{for $\varepsilon\leqslant-\frac{1}{2}$},\\
\displaystyle\vphantom{\int\limits_-}\int_{-2\varepsilon}^1 {\mathrm d}y\,y^{-1/2}\left(1-y\right)^{(N-3)/2}\left(2\varepsilon+y\right)^{(N-3)/2} & \mbox{for $-\frac{1}{2}<\varepsilon\leqslant0$},\\
\displaystyle\int_0^1 {\mathrm d}y\,y^{-1/2}\left(1-y\right)^{(N-3)/2}\left(2\varepsilon+y\right)^{(N-3)/2} & \mbox{for $0<\varepsilon$},
}
\end{equation}
we observe that the ($0<\varepsilon$)-case is, apart from some prefactor, the standard form of the integral representation of the Gauss hypergeometric function $\Fhyper$ \cite{MaObSo}, whereas the integral for $-\frac{1}{2}<\varepsilon\leqslant0$ can be transformed into such a standard form by a substitution of variables $y\to1-y(1+2\varepsilon)$. Then, in terms of Gauss hypergeometric functions and Gamma-functions $\Gamma$, we obtain
\begin{equation}\label{eq:Omega_final}
\fl\Omega_N(\varepsilon) \propto
\cases{
0 & \mbox{for $\varepsilon\leqslant-\frac{1}{2}$},\\
{\Gamma\left(\case{N-1}{2}\right)\Gamma\left(\case{N}{2}\right) \left(1+2\varepsilon\right)^{N-2} \Fhyper\left(\case{1}{2},\case{N-1}{2},N-1;1+2\varepsilon\right)} & \mbox{for $-\frac{1}{2}<\varepsilon\leqslant0$},\\
{\sqrt{\pi}\,\Gamma\left(N-1\right) \left(2\varepsilon\right)^{(N-3)/2} \Fhyper\left(\case{1}{2},\case{3-N}{2},\case{N}{2};-\case{1}{2\varepsilon}\right)} & \mbox{for $0<\varepsilon$}.
}
\end{equation}
It follows from the properties of these functions that, in the interior of its domain $[-\frac{1}{2},\infty)$, the density of states $\Omega_N(\varepsilon)$ has precisely one nonanalytic point at $\varepsilon=0$ for all system sizes $N\geqslant2$. 

In order to compute the thermodynamic limit of the microcanonical entropy,
\begin{equation}
s(\varepsilon)=\lim_{N\to\infty}\frac{1}{N}\ln\Omega_N(\varepsilon),
\end{equation}
we perform a substitution of variables $y=x^2$ in \eref{eq:Omega6}, obtaining
\begin{equation}
\Omega_N(\varepsilon) \propto \int_{\max\{0,-2\varepsilon\}}^1 {\mathrm d}x\,\left[\left(1-x^2\right)\left(2\varepsilon+x^2\right)\right]^{(N-3)/2}
\end{equation}
for $\varepsilon>-\frac{1}{2}$. The large-$N$-limit of this integral can be evaluated by Laplace's method \cite{BenOrs}, yielding, to leading order, the maximum of its integrand on the domain of integration. Hence, omitting a physically irrelevant additive constant, we can write
\begin{equation}
s(\varepsilon)=\lim_{N\to\infty}\frac{1}{N}\ln\max_{x\in\left[\max\{0,-2\varepsilon\},1\right]}\left[\left(1-x^2\right)\left(2\varepsilon+x^2\right)\right]^{(N-3)/2}.
\end{equation}
The maximum in this expression is attained at $x=\sqrt{\frac{1}{2}-\varepsilon}$ for $-\frac{1}{2}<\varepsilon\leqslant\frac{1}{2}$, and at $x=0$ for $\varepsilon>\frac{1}{2}$, and we obtain  
\begin{equation}
s(\varepsilon)=
\cases{
\ln\left[\case{1}{2}(1+2\varepsilon)\right] & \mbox{for $\varepsilon\leqslant\frac{1}{2}$},\\
\case{1}{2}\ln(2\varepsilon) & \mbox{for $\varepsilon > \frac{1}{2}$},
}
\end{equation}
for the microcanonical entropy in the thermodynamic limit. The same result can be obtained directly from equation~ (\ref{eq:Omega_final}) by making use of the asymptotic expressions for Gauss hypergeometric functions $\Fhyper$ \cite{OldeDaalhuis:03}. The singularity which was present at finite $N$ for $\varepsilon=0$ has now moved to $\varepsilon=\frac{1}{2}$. This was previously noted and discussed in reference \cite{CasettiKastnerPRL06} where a short account of the solution of the mean-field spherical model with kinetic energy was given.\\

\bibliographystyle{unsrt}
\bibliography{partial}

\end{document}